\documentclass{article}
\usepackage{amssymb} \usepackage{amsfonts}
\usepackage{graphicx}
\newcommand{\dd}{\stackrel{^-}{}\!\!\!\!\!\!d\,}

\begin{document}



\title{Nuclear matter superfluidity in an effective hadronic field model with excluded
volume corrections}

\author{ R. M.  Aguirre\\
Departamento de F\'{\i}sica, Facultad de Ciencias Exactas,\\
Universidad Nacional de La Plata.\\
C. C. 67 (1900) La Plata, Argentina\\
and IFLP, CCT La Plata, CONICET.}
\date{ }

\maketitle

\begin{abstract}
Properties of the $^1S_0$ superfluid phase are studied for
symmetric nuclear matter at finite temperature. It is described
within a covariant hadronic field model, of the $\sigma -\omega$
type, with addition of density dependent correlations simulating
effects due to finite extension of nucleons. The model is solved
in a selfconsistent Hartree-Bogoliubov approach, assuming
instantaneous interactions in the superfluid phase. A comparison
with the results obtained from several hadronic field models is
made. Main characteristics of our description of the superfluid
gap are in qualitative agreement with some studies using
microscopic potentials, although further refinements could improve
its performance.\end{abstract}

\section{Introduction}

Superfluid states in the nuclear environment have been extensively
studied as they have a significative role in several physical
processes, such as the structure of nuclei out the
stability valley and the cooling dynamics of proto-neutron stars.\\
A variety of models and approximations have been used for this
purpose
\cite{TANIGAWA,SERRA,LONG_nt0812,MUTHER_prc72,ZUO_prc66,GANDOLFI_prl101,CAO_prc74,MATERA,KUCHAREK,CHEN_plb585,LiSunMeng,MATSUZAKI3,DEAN_rmp,MATSUZAKI0,MATSUZAKI1,MATSUZAKI2},
mainly non-relativistic potentials or effective forces, such as the
density-dependent Skyrme or Gogny ones. It is not unusual a mixed
treatment that combines both schemes in order to simplify
involved calculations \cite{TANIGAWA,SERRA}.  \\
Approximately two decades ago a covariant model of the field theory
of hadrons, generally known as Quantum Hadro-Dynamics (QHD)
\cite{QHD1a,QHD1b}, was used for the first time to study nuclear
matter superfluidity \cite{KUCHAREK}. There are several reasons to
use this theoretical framework to deal with nuclear superfluidity,
in first place there are practical reasons, some self-consistent
calculations are more simply stated and easily solved. This property
gave rise to a version of QHD, known as Density Dependent Hadron
Field Theory \cite{HOFMANN_prc64} which casts Dirac-Brueckner
outputs in the QHD language. In second place one must consider field
theory as a more adequate tool to make contact with the fundamental
theory of strong interactions and the fact that covariant
formulations are desirable for astrophysical applications, among
others conceptual reasons. Furthermore, the formalism has the
versatility to include  vacuum effects and
finite renormalizations in a coherent way \cite{MATSUZAKI0,MATSUZAKI1}.\\
Since QHD models are formulated as a many-body theory, one of its
basic premises is the reproduction of the nuclear matter
saturation properties. This can be fulfilled with a few adjustable
parameters and even at the lowest order of approximation. Once the
free-parameters have been fixed, the QHD model has a noticeable
prediction power.\\
Although a wide spreading of numerical results for the superfluid
phase in infinite nuclear matter can be found in the literature,
there is a qualitative agreement that the superfluid gap should
not exceed 3 MeV, and it should vanish for densities around the
saturation density \cite{DEAN_rmp}. Unfortunately, the findings of
\cite{KUCHAREK} for superfluid nuclear matter in the $^1S_0$ phase
do not agree with these expectations. A maximum
$\Delta_{max}(p_F)\approx 10$ MeV was found there, and only an
unphysical reduction of 15\% in the omega-meson mass yield results
comparable with currently accepted values.\\
This situation has not been changed substantially after the
evaluation of different corrections into the original scheme.
However, the good properties of this treatment have motivated
mixed descriptions combining QHD models and conventional
potentials \cite{TANIGAWA,SERRA,LONG_nt0812}.\\
More recently it was claimed that a coherent inclusion of meson
proper self-energy \cite{CHEN_plb585}, or adjustable quenching
factor \cite{LiSunMeng}, could bring numerical calculations to the
likely values . Further studies about the effect of corrections of
the meson propagators and the influence of low density instabilities
on the superfluid gap can be found for instance in \cite{MATSUZAKI3}.\\
It was stressed in reference \cite{KUCHAREK} that the high momenta
behavior of the repulsive potential has a crucial role in the
exceedingly large values obtained for $\Delta_{max}(p_F)$. It must
be noticed that the relative strength of repulsive and attractive
contributions have been calibrated for momenta below the Fermi
surface  in order to produce the saturation mechanism. Therefore
it would be desirable a pairing potential which preserves the
relative strength of its components in the Fermi sphere, but
having a repulsive component decreasing as faster as the
attractive one in the high momenta domain.\\

Taking these facts into account we try in the present work, to
obtain an effective and concise model able to deal with the
superfluid phase of nuclear matter.  For this purpose we use and
compare several models of relativistic nuclear fields interacting
through scalar and vector mesons. As a first approach, we reduce to
its minimal expression the complexity of the nuclear interaction,
but further refinements can be considered. We introduce a
characteristic length scale in the effective interaction, which
could  eventually be traced back to the confining mechanism of the
fundamental theory of strong interactions. A similar approach was
applied in the past to describe heavy ion collisions
\cite{StoGre,Cley1,Cley2,Cley3,Cley4} as well as nuclear matter
properties
\cite{KAGIYAMA1,KAGIYAMA2,KAGIYAMA3,SINGH1,SINGH2,SINGH3}. It has
been known as finite volume correction since it takes into account
the spread of the nucleon localization. The relevance of this effect
upon the evaluation of some bulk properties of the nuclear matter
has been
stressed long time ago \cite{KAPUSTA}.\\

 In the next section we present the theoretical deduction of the
general gap equation at finite temperature in a context of
Landau-Fermi liquid.  The last part of this section is devoted to
the theoretical deduction of the self-consistent expression
defining the superfluid gap for the $^1S_0$ phase within the QHD
framework. In section 3 we show the numerical results and discuss
them. Finally, conclusions are drawn in section 4.

\section{The Formalism} \setcounter{equation}{0}

\subsection{Superfluid states in a Landau-Fermi liquid}

The different superfluid phases in a fermionic system can be
described in a general and compact way within the formalism of the
Fermi liquid, using for instance the formalism of reference
\cite{AKHIEZER}. It is assumed that the low-lying excitations of
the system are represented in terms of quasi-particles and,
circumstantially, collective modes. \\
We use the notation $f_\alpha$ for the equilibrium distribution
function of a quasi-particle state, where the label $\alpha$
comprise spin, isospin, and momentum quantum numbers.  The
fermionic contribution to every conserved quantity, such as
particle number and energy,
can be expressed in terms of a summation over $f_\alpha$.\\
In the following we will be interested in nucleon pairs coupled to
singlet spin and triplet isospin, so that an anomalous
distribution function $g_\beta$ and a energy gap $\Delta_\beta$
are introduced. According to \cite{AKHIEZER}, we make the
decomposition
\begin{eqnarray}
g_\beta&=&g_k(p)\left[ \tau^{(k)} \tau^{(2)}\right]_{a\, b} \sigma^{(2)}_{s\,s'},  \label{GDIST}\\
\Delta_\beta&=&\Delta_k(p)\left[ \tau^{(k)} \tau^{(2)}\right]_{a\,
b} \sigma^{(2)}_{s\,s'}, \label{GAPDECOMP}
\end{eqnarray}
\noindent
 where momentum, isospin $(a,b)$, and spin $(s,s')$ dependencies has been
clearly distinguished. Here $\tau_k, \, \sigma_k, \,k=1,2,3$
stands for the Pauli matrices for isospin and spin,
respectively.\\
On the other hand, the normal phase is filled with
 quasi-particle states in a isospin duplet described similarly by  $f_\beta=\delta_{s s'}f_{a
b}(p)$, with
\[ f_{a\,b}(p)=f_0(p) \delta_{a\,
b}+f_3(p)\tau^{(3)}_{a\, b}. \label{FDIST}
\]
Accordingly, the quasi-particle spectra is assumed in
matrix form $\varepsilon(p)=\varepsilon_0(p)+\varepsilon_3(p)\, \tau_3$.\\

 In the Landau theory of Fermi liquids, the energy of the system
$E$ is considered as a functional of the distribution functions
$f_\alpha,\, g_\beta$. First variations respect to them give the
quasi-particle and gap spectra matrix components
\begin{equation}
\varepsilon_{jk}(p)=\frac{\delta E}{\delta f_{jk}(p)},\;\;
\Delta_{jk}(p)=\frac{\delta E}{\delta g_{jk}^\dag(p)}.
\label{SPECTRA}
\end{equation}

Since $f$ and $g$ are itself functions of $\varepsilon,\, \Delta$,
the equations above are self-consistent relations.\\
Within the block diagonalization procedure of \cite{AKHIEZER} the
distribution functions are written
\begin{eqnarray}
f&=&K\,n+X(1-n^t)X^\dag K, \label{FDIST_d} \\
g&=&(K n X)^t+K (1-n) X,  \label{GDIST_d}\\
n&=&\left[1+e^{\beta (\xi-X \Delta^\dag)} \right]^{-1}, \label{NDIST}\\
K&=&\left(1+ X X^\dag \right)^{-1},
\end{eqnarray}

\noindent here $t$ indicates matrix transposition, $\beta=1/kT$,
$\xi=\varepsilon+\mu$, and the diagonal matrix
$\mu=$diag$(\mu_1,\mu_2)$ collects the proton (1) and nucleon (2)
chemical potentials. The unknown matrix $X=X_j \tau_j \tau_2
\sigma_2$ satisfies the condition
\[ \xi X+ X \xi^T+\Delta-X \Delta^\dag X=0
\]

We have solved this system of equations for symmetric nuclear
matter coupled to $T_z=0$. We obtained
\begin{eqnarray}
&&X_i=\Delta_i=g_i=0,\, \mbox{for  } i=1,2, \, X_3=\frac{\varepsilon_0-\mu_0\pm E_\Delta}{\Delta_3^\ast} \nonumber\\
&&g_3(p)=-\frac{\Delta_3}{2 E_\Delta} \tanh(\beta
E_\Delta/2),\,f_0=\frac{1}{2}\left[1-\frac{\varepsilon_0-\mu_0}{E_\Delta} \tanh(\beta E_\Delta/2)\right],  \nonumber\\
&& f_3=0,\, E_\Delta=\sqrt{\Delta_3^2+(\varepsilon_0-\mu_0)^2};
\nonumber
\end{eqnarray}
 in the first line $\mu_0$  stands for either the proton or the neutron chemical potential.\\
 The particle density of protons (k=1) or neutrons (k=2)
 is given by\\
 \[ n_k= \int \frac{d^3p}{(2 \pi)^3} \left[f_0(p)+\tau_{kk}^{(3)}
 f_3(p)\right]\nonumber
 \]
these equations are used to relate the chemical potentials to the
conserved isospin and baryonic number density.\\
In the next subsection we show the model which provides the
quasi-particle interaction and spectra.

\subsection{Hadronic models}

Models of the nuclear interaction, formulated in the covariant field
theory, have been widely used in the study of the dynamics and
structure of infinite matter as well as finite nuclei. Since the
pioneering work of references \cite{QHD1a,QHD1b}, the simple
$\sigma-\omega$ model has grown in different directions and it was
completed in order to cover a multitude of manifestations of the nuclear force.\\
In particular the subject of the nuclear superfluidity was first
treated within this context in reference \cite{KUCHAREK}, by using
the original $\sigma-\omega$ model plus a pseudo-scalar pion
interaction. The scheme of approximation used there consisted in a
mean field treatment of the meson and nucleon fields, a Gorkov
factorization of the pairing interaction and a instantaneous
assumption which allows a time-independent resolution of the gap
equation. Subsequently, this procedure was extended to consider
the effect of vacuum, and the variation of the in-medium
meson properties \cite{MATSUZAKI0,MATSUZAKI1}. \\
Most of these studies agree in a excessively large value for the
gap in infinite nuclear matter. The realization of a correlated
state of two nucleons is a consequence of the equilibrium between
a repulsive and an attractive component of the pairing potential,
originating in the exchange of virtual $\omega$-mesons and
$\sigma$-mesons respectively. This mechanism is also found in the
binding energy of nuclear matter.\\
The slow decrease of the repulsive potential as a function of the
transferred momenta, has been pointed out as the main cause of the
mismatch. This situation can not be modified without a substantial
 redefinition of the couplings, which should lead to a destruction of the
saturation mechanism.\\

In this work we adopt the simplest version of nucleons interacting
through $\sigma$ and $\omega$ mesons, whose lagrangian density is\\
\[ {\cal L}= \bar{\Psi}^a\left(i \not \! \partial -M_a +g_s \sigma- g_w
\not \! \omega \right) \Psi^a + \frac{1}{2} (\partial^\mu \sigma
\partial_\mu \sigma - m_s^2 \sigma^2)
-\frac{1}{4}
 F^{\mu \nu} F_{\mu \nu} + \frac{1}{2} m_w^2
\omega^2 \nonumber
\]

\noindent where summation over the repeated isospin index $a$ is
assumed, $F_{\mu \nu}=\partial_\mu \omega_\nu-\partial_\nu
\omega_\mu$ and $g_s, g_w$ are adimensional coupling constants.
The equations of motion of the classical fields are
\begin{eqnarray}
\left(i \not \! \partial -M_a +g_s \sigma- g_w \not \! \omega
\right)
\Psi^a&=&0, \label{NUCLEONEQ} \\
\left(\Box+m_s^2 \right) \sigma-g_s \bar{\Psi}^a \Psi^a&=&0\label{SIGMA} \\
\partial_\nu F^\nu_\mu+m_w^2 \omega_\mu -g_w \bar{\Psi}^a \gamma_\mu
\Psi^a&=&0
  \label{OMEGA}
\end{eqnarray}

Denoting by $D(x,y), \, D_{\mu \nu}(x,y)$ the propagators of
scalar and vector-meson fields, the eqs. (\ref{SIGMA}) and
(\ref{OMEGA}) can be formally solved as
\begin{eqnarray}
 \sigma(x)&=&g_s\int \dd^4y D(x,y) \Psi^a(y) \Psi^a(y), \\
 \omega_\mu(x)&=&g_w\int \dd^4y D_{\mu \nu}(x,y) \Psi^a(y) \gamma^\nu \Psi^a
(y) \label{MESONS}
\end{eqnarray}

The Hamiltonian density $\mathcal{H}$ is given by the canonical
procedure, and the energy density of the system $\mathcal{E}$ is
evaluated by taking its expectation value

\begin{equation}
\mathcal{E}=<\bar{\Psi}^a i \gamma_0 \partial_0 \Psi^a+\partial_0
\sigma \partial_0 \sigma -F_{0
\mu}\partial_0\omega^\mu+\frac{1}{2}\sigma
(\Box+m_s^2)\sigma-\frac{1}{2} \omega_\nu (\partial_\mu F^{\mu
\nu}+m_w^2 \omega^\nu)> \label{E1}
\end{equation}

Under the hypothesis of static meson fields the second and third terms vanish.\\
Inserting the meson equations of motion (\ref{SIGMA}),
(\ref{OMEGA}) together with the Eq. (\ref{MESONS}) into
(\ref{E1}), we obtain
\begin{eqnarray}
\mathcal{E}(x)&=&<\bar{\Psi}^a(x) i \gamma_0 \partial_0
\Psi^a(x)>+\frac{g_s^2}{2}\int \dd^4y D(x,y) <\Psi^a_\alpha(x)
\Psi^a_\alpha(x)\Psi^b_\beta(y)
\Psi^b_\beta(y)>\nonumber \\
&&-\frac{g_w^2}{2}\int \dd^4y  D_{\mu \nu}(x,y) \gamma^\mu_{\alpha
\alpha'} \gamma^\nu_{\beta \beta'}) <\Psi^a_\alpha(x)
\Psi^a_{\alpha'}(x)\Psi^b_\beta(y) \Psi^b_{\beta'}(y)>
\label{DENSE}
\end{eqnarray}

At this point we introduce an expansion of the nucleon fields in
terms of quasi-particle
 creation and annihilation operators, similar to that of a free
 field
\begin{eqnarray}
\Psi^a_\alpha(x)&=&\int \dd^3p
\frac{M^\ast_a}{E_a(p)}\left[b^a_s(p) u^s_\alpha(p) e^{-i P_a
 x}+ d^{\dag \, a }_s(p) v^s_\alpha(p) e^{i P_a  x}\right],
\label{PSIQUANT} \\
\bar{\Psi}^a_\alpha(x)&=&\int \dd^3p
\frac{M^\ast_a}{E_a(p)}\left[b^{\dag \, a }_s(p)
\bar{u}^s_\alpha(p) e^{i P_a x}+d^a_s(p) \bar{v}^s_\alpha(p) e^{-i
P_a  x}\right],
\label{PSIBARQUANT} \\
u_a^s(p)&=&\frac{\not \! P_a+M^\ast_a}{\sqrt{2 M^\ast_a(M^\ast_a
+E_a (p))}}\;u^s(0),\nonumber \\
v_a^s(p)&=&\frac{\not \! P_a+M^\ast_a}{\sqrt{2 M^\ast_a (M^\ast_a
+E_a (p))}}\;u^s(0), \nonumber
\end{eqnarray}
\noindent but creation and annihilation operators are referred to
the lowest energy state of correlated nucleons. We have used the
quasi-particle properties $M^\ast_a=M-g_s\, s$, $E_a(p)=
\sqrt{M^{\ast\,2}_a+p^2}, P_a=(E_a(p)+g_w w, \mathbf{p})$. Due to
the isospin invariance of the interaction proton and neutron
properties are actually independent.  Furthermore, as we are
interested in isospin symmetric matter proton and neutron are
indistinguishable and the isospin index $a$ becomes superfluous.
From now on we will omit
it, and a degeneracy factor 2 will be included when necessary.\\
The quantities $s,\, w$ stand for the mean values of the $\sigma$
and $\omega$ meson fields in homogeneous,
 isotropic matter. As usual, they can be deduced from Eqs. (\ref{SIGMA}), (\ref{OMEGA})
by neglecting derivatives and taking expectation values of the
fermionic bilinears
$s=2\, g_s<\bar{\Psi}\Psi>/m_s^2$, $w=2\, g_w<\bar{\Psi}\gamma_0 \Psi>/m_w^2$.\\
As a part of the approximation we neglect in Eq. (\ref{DENSE}) the
contribution of particle-antiparticle or antiparticle-antiparticle
terms. We define the equilibrium
 distribution functions for the normal and superfluid phases
\begin{eqnarray}
f_{ss'}(p,k)&=&\frac{M^\ast}{\sqrt{E(p) E(k)}}<b^{\dag}_{s'}(k) b_s(p)>,\\
g_{ss'}(p,k)&=&\frac{M^\ast}{\sqrt{E(p) E(k)}}
<b_{s'}(k) b_s(p)>,\\
g^{\dag}_{ss'}(p,k)&=&\frac{M^\ast}{\sqrt{E(p) E(k)}} <b^{\dag
}_{s'}(k) b^{\dag}_s(p)>.
\end{eqnarray}
It must been taken into account that $f(k,p)\propto (2 \pi)^3
\delta^3(p-k)$, whereas pairs of nucleons are assumed to couple to
zero momentum, so that $g(k,p)\propto(2 \pi)^3\,
\delta^3(p+k) $.\\

Within the mean field approach, corrections to the meson
propagation in the nuclear environment are dismissed, although a
random phase approximation could be considered, as in
\cite{CHEN_plb585}. In the first case, we obtain
\begin{eqnarray}
D(z)&=&\int \dd^4q D(q) e^{-i q z} \label{FOURIERT} \\
D(q)&=&-(q_\lambda q^\lambda-m_s^2+i\varepsilon)^{-1}, \label{SIGMAPRO} \\
D_{\mu \nu}(q)&=&-(g_{\mu \nu}-q_\mu q_\nu/m_w^2)/(q_\lambda
q^\lambda-m_w^2+i\varepsilon),  \label{OMEGAPRO}
\end{eqnarray} \noindent
The term proportional to $q_\mu$ in (\ref{OMEGAPRO}), produce zero
contribution in integrals combining $D_{\mu \nu}$
with nucleon fields, because of the conservation of the baryonic current.\\
We are interested in static homogeneous matter, therefore $z=x-x'$
in Eq. (\ref{FOURIERT}).  If we neglect time retardation in the
meson propagation, {\it i. e.} $0=x_0-x'_0$, then terms containing
$q_0$ are absent in Eqs. (\ref{SIGMAPRO}), (\ref{OMEGAPRO}) as
well as in the exponential of the Fourier transform shown lines above.\\
All these elements together in Eq. (\ref{DENSE}) produce
\begin{eqnarray}
\mathcal{E}&=&\mathcal{E}_{MFA}+\mathcal{E}_F+\mathcal{E}_S \nonumber \\
\mathcal{E}_{MFA}&=&4 \int \dd^3p \,\varepsilon(p)
f(p)+2\left(\frac{2\, g_s}{m_s}\int\dd^3p \frac{M^\ast}{E(p)}
f(p)\right)^2\nonumber \\&&-2\left(\frac{2\, g_w}{m_w}\int\dd^3p f(p)\right)^2  \\
\mathcal{E}_F&=&2\, g_s^2\int \dd^3p\, \dd^3q \frac{f(p) f(q)}{ 2
E(p)E(q)} \frac{M^{\ast\,2} +E(p) E(q)-\mathbf{p} \cdot
\mathbf{q}}
{\left[\varepsilon(p)-\varepsilon(q)\right]^2-\mid \mathbf{p}-\mathbf{q} \mid^2-m_s^2+i \varepsilon}\nonumber\\
&&- 2\, g_w^2\int \dd^3p\, \dd^3q \frac{f(p) f(q)}{ E(p)E(q)}
\frac{2 M^{\ast\,2}-E(p) E(q)+\mathbf{p} \cdot \mathbf{q}}
{\left[\varepsilon(p)-\varepsilon(q)\right]^2-\mid
\mathbf{p}-\mathbf{q} \mid^2-m_w^2+i \varepsilon}
\label{EFOCK}\\
\mathcal{E}_S&=&-g_s^2\int \dd^3p\, \dd^3q \,\frac{g^{\dag}(p)
g(q) \Lambda_s(p,q)}
{\left[\varepsilon(p)-\varepsilon(q)\right]^2-\mid
\mathbf{p}-\mathbf{q} \mid^2-m_s^2+i \varepsilon}\;\eta(p,q)\nonumber \\
&&+g_w^2\int \dd^3p\, \dd^3q \,\frac{g^{\dag }(p) g(q)
\Lambda_w(p,q)} {\left[\varepsilon(p)-\varepsilon(q)\right]^2-\mid
\mathbf{p}-\mathbf{q} \mid^2-m_s^2+i \varepsilon}\;\eta(p,q).
\label{ESUP}
\end{eqnarray}

Here we have separated mean field ($MFA$), Fock ($F$) and
superfluid ($S$) contributions. Certain integrals appearing in
(\ref{DENSE}) vanish because of the isotropy of infinite matter.
The following notation is used
\begin{eqnarray}
\Lambda_s(p,q)&=&\frac{X^2(p,q)-2 X(p,q)\mathbf{p}
\cdot \mathbf{q}+p^2q^2}{E(p) E(q) X(p,q)}, \nonumber \\
\Lambda_w(p,q)&=&\Lambda_s(p,q)+ \frac{3 X(p,p)q^2+3 X(q,q)p^2-2
X(p,q)\mathbf{p} \cdot \mathbf{q}+p^2 q^2}{E(p) E(q) X(p,q)}, \nonumber\\
\eta(p,q)&=&\exp\left(-i 2 x_0[\varepsilon
(q)-\varepsilon(p)]\right),\nonumber
\end{eqnarray}
\noindent with $X(p,q)=(M^\ast+E(p))(M^\ast+E(q))$.\\

The factor $\mathbf{p} \cdot \mathbf{q}= p \,q \nu$ entering in
these expressions can be used for integration respect to $\nu$.
For this purpose it is useful the distribution identity\\
\[ \frac{1}{z+i\varepsilon}=PV\left( \frac{1}{z}\right)- i \pi \delta(z)\nonumber\]
Denominators in the integrands of Eqs. (\ref{ESUP}), (\ref{EFOCK})
come from the meson propagators, in particular the combinations
$\varepsilon(q)-\varepsilon(p)$ come from its $q_0$ dependence. In
the next step we apply the instantaneous approximation
\cite{KUCHAREK}, which result in the elimination of
all these combinations.\\
Within this approach Eq. (\ref{ESUP}) can be
re-written\\
\begin{eqnarray}
\mathcal{E}_S&=&-g_s^2\int \dd^3p\, \dd^3q  \frac{g(q) g^{\dag}(p)
}{E(p)E(q)}\nonumber \\
&+&g_s^2 \int \dd^3p\, \dd^3q \,g(q)\, g^{\dag}(p)\frac{X^2(p,q) +
p^2q^2 -X(p,q)(p^2+q^2+m_s^2)}{4 p\, q\, E(p)E(q)X(p,q)}
\ln\left(\frac{(p-q)^2+m_s^2}{(p+q)^2+m_s^2}
\right) \nonumber \\
&-&g_w^2 \int \dd^3p\, \dd^3q \,g(q)\, g^{\dag}(p)\frac{X^2(p,q) +
p^2q^2+3 X(p,p) q^2+3 X(q,q) p^2}{4 p\, q\, E(p)E(q)X(p,q)}
\ln\left(\frac{(p-q)^2+m_w^2}{(p+q)^2+m_w^2} \right) \nonumber
\end{eqnarray}
From now on, we use only the MFA and neglect the Fock term.\\
 The  superfluid gap can be
determined by using Eqs. (\ref{GAPDECOMP}) and (\ref{SPECTRA})
\begin{eqnarray}\Delta_3(p)&=&-\frac{g_s^2}{8}\int \dd^3q
\frac{g_3(q)}{E(p)E(q)} + \frac{g_s^2}{16}\int \dd^3q
\,g_3(q)\,\ln\left(\frac{(p-q)^2+m_s^2}{(p+q)^2+m_s^2}
\right)\frac{4 M^2- (E(p)- E(q))^2 -m_s^2}{p \,q
E(p)E(q)}\nonumber \\
&& -\frac{g_w^2}{8}\int \dd^3q
\,g_3(q)\,\ln\left(\frac{(p-q)^2+m_w^2}{(p+q)^2+m_w^2}
\right)\frac{2 E(p)E(q)-M^2}{p\, q E(p)E(q)} \label{SNMgap}
\end{eqnarray}

Finally it must be stressed that in applying Eq.(\ref{SPECTRA})
for evaluating the quasi-particle spectra, the effective mass
$M^\ast$ must be considered a functional of the distribution
function $f$.

\subsection{Effects of the spatial extension of nucleons}

Standard field theory considers physical particles as
structureless, point-like objects. This could be a serious
shortcoming when composed states in a dense medium are described.
It is well known that, for instance, the energy and charge density
of a soliton spreads over a finite range of space \cite{ZAHED}.
Therefore it is legitimate to assign a intrinsical length scale to
nucleons immersed in a dense environment.\\
This was the argument supporting many phenomenological studies of
the nuclear interaction \cite{SINGH1,SINGH2,SINGH3,KAPUSTA}. From a
practical point of view, one can consider  $N_a$ fermions of class
$a$ distributed over a finite volume $V$, then the available space
for quantization is $V'=V-\sum_a N^a v_a$, where $v_a$ is the
spatial extension of this state. But the canonical procedure uses
the full volume $V$, this situation is corrected by introducing a
correction factor $\sqrt{V'/V}$ in the second quantization of
fields. This, in turn, modifies fermion bilinears like particle
number $n_a=<\bar{\Psi}_a\Psi_a>$ and normal energy density by a
factor $\Theta=V'/V=1-\sum_b n^b v_b$. The superfluid energy density
requires a correction $\Theta^2$, which is transferred to
the formulae of the gap function.\\
The effective volume inaccessible for other particles due to the
presence of a spherical object of radius $R_a$ is
\[
v_a=\alpha \frac{4 \pi}{3} {R_a}^3, \nonumber
\]
The parameter $R_a$ can be understood in a simple minded model as
the geometrical size of a particle. Actually it introduces into the
model a characteristic scale of the strong interaction, i. e. the
spatial spreading of a bounded state of quarks and gluons. The value
selected for this radius must be compatible with similar lengths
adopted in hybrid models of the nuclear interaction, see for
instance \cite{THOMAS}. Returning to the schematic picture, the
factor $\alpha$ takes account of the fact that the inaccessible
volume exceeds the actual size of each particle and it depends on
the spatial arrangement adopted by the collection of objects. The
minimal volume configuration for identical particles corresponds to
a face centered cubic arrangement. In such a case is
$\alpha=3\sqrt{2}/{\pi}$, which is the value adopted in the present
calculations.\\
It is worthwhile to mention that the normalization of the nucleon
field with an excluded volume coefficient is not equivalent to the
introduction of a hard-core potential. The normalized nucleon
field interacts dynamically with the meson fields, both scalar and
vector. The in-medium properties of protons and neutrons, as well
as the meson fields configuration arise simultaneously from this
interaction. The sigma meson gives rise to the attractive channel
of the nuclear force, whereas the omega meson is responsible for
the repulsive component. Therefore, the proposed normalization
affects both attractive and repulsive channels. This fact is
evident from Eq. (\ref{SIGMA}) and the discussion given above. It
should be clear that the mean-field value of the sigma meson is
strongly affected by the normalization of the nucleons. Moreover,
the relation is highly non-linear. Furthermore, the treatment of
the mesons is not symmetrical since the omega meson is coupled to
a conserved charge. In consequence the omega meson mean field
value is completely determined by the conserved baryonic density.
The sigma meson mean field value, instead, come forth the hadronic
dynamics.\\

Since $\Theta$ introduces an explicit dependence upon the baryonic
densities, the quasi-particle energy $\varepsilon$, see Eq.
(\ref{SPECTRA}),
gets an extra term\\
\[
\varepsilon_a(p)=\sqrt{p^2+M^{\ast\, 2}_a}+g_w w+\frac{2
v_a}{3}\int \dd^3q \frac{q^2\, f^a(q)}{E_a(q)},
\]
\noindent it must be noticed that the additional term depends on
density and temperature, but not on the momentum.

\section{Results and Discussion}
\setcounter{equation}{0}

The model of nucleons and mesons with finite volume corrections
 has several parameters, we used $M=940$ MeV for the mass of the
degenerate nucleons, and $m_w=783$ MeV for the omega-meson mass.
The sigma-meson mass has been fixed at $m_s=520$ MeV, in agreement
with \cite{KUCHAREK}. A discussion about variation of $m_s$ can be
found in \cite{KUCHAREK}.\\
The length scale $R_a$ is not determined by the model, therefore
we consider it as a constant value ranging
between 0.5 fm and 0.9 fm. \\
The coupling constants $g_s,\, g_w$ are fixed in order to
reproduce the saturation properties of symmetric nuclear matter,
the binding energy $E_B=-16$ MeV and a saturation density
corresponding to the Fermi momentum $p_F=1.42 \, \mbox{fm}^{-1}$
\cite{KUCHAREK,LiSunMeng}.

 To have a look of the performance of QHD models in
describing the nuclear superfluidity, we have examined two different
models of the nucleon-meson interaction. In first place we study the
density dependent coupling model of \cite{HOFMANN_prc64}, which
translate Dirac-Bruckner calculations with Bonn A potential into the
covariant field theory. We have used the rational function
parametrization given there for the couplings, obtaining a maximum
value $\Delta_{max}(p_F)\simeq 15$ MeV  at $n/n_0=0.25$. As another
QHD example we take the non-linear meson-nucleon model of
\cite{ZIMANYI}, it predicts a lower compressibility and higher
effective mass in the MFA than the $\sigma-\omega$ model
\cite{QHD1a,QHD1b} does. In our calculations we get
$\Delta_{max}(p_F)=10.6$ MeV at $n/n_0\simeq
0.1$ .\\
\begin{table}[pt]\vspace{-0.25cm}\label{TABLE}
 {\begin{tabular}{ccc}
$R$[fm]&$\Delta_{max}^{FEC}/\Delta_{max}^{NFEC}$ &$g_w/g_s$\\
\hline
0.6  & 0.68  &1.222\\
0.62&  0.64  &1.209\\
0.64&  0.61  &1.193\\
0.66&  0.59  &1.171\\
0.68&  0.58  &1.141\\
0.70  & 0.56  &1.095\\
0.71 & 0.57  &1.062\\
0.72 & 0.60  &1.018
\end{tabular}}
\caption{The gap function on the Fermi surface and the quotient of
coupling constants for several values of the characteristic
nucleon size $R$.}\end{table}

As the next step we consider the effects of the finite extension
of nucleons, taking $R$ as an adjustable parameter. The results
are summarized in Table 1, where the treatments regarding the
finite extension of nucleons (FEC) or neglecting it (NFEC) have
been distinguished. There is a sensible reduction of
$\Delta_{max}$, between 30\% and 40 \%, respect to the previous
calculations. A non-monotonous  dependence on $R$ is obtained, the
lowest value $\Delta_{max}=5.5$ MeV is reached for $R=0.7$ fm at
$p_F=0.8$ fm$^{-1}$. \\
Choosing $R=0.7\, \mbox{fm}^{-1}$ for subsequent calculations, we
compare the momentum dependence of the gap function for a fixed
density $n/n_0=0.25$. Results are shown in Fig.1. The square
points show the magnitude of the gap at the Fermi surface. A
comparison of the two cases NFEC and FEC, shows that the latter
yields the lowest absolute value for $q < 1.6$ GeV. A reduction of
almost $40 \%$ is registered at the Fermi surface in the FEC case.

\begin{figure}[hb]
\centering 
\includegraphics[height=0.4\textheight]{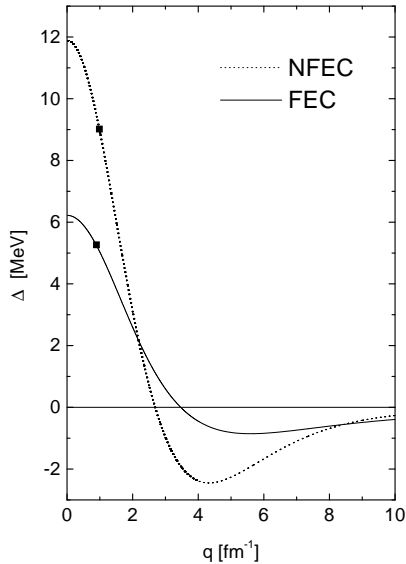}
\vspace{-0.5cm} \caption{The gap energy as a function of the
 momentum, square symbols indicate its value at the Fermi surface.}\label{FIG1}
\end{figure}

\begin{figure}[ht]
\centering
\includegraphics[height=0.4\textheight]{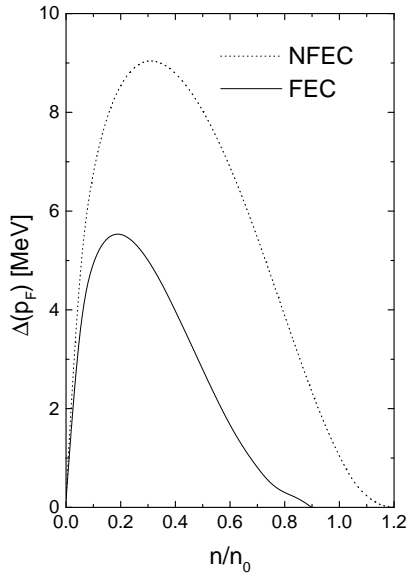}
\vspace{-0.5cm} \caption{The energy gap evaluated on the Fermi
surface in terms
 of the particle density. Results including Finite Extension Corrections (FEC)
 or not (NFEC).}\label{FIG2}
\end{figure}

\begin{figure}[ht]
\centering
\includegraphics[height=0.45\textheight]{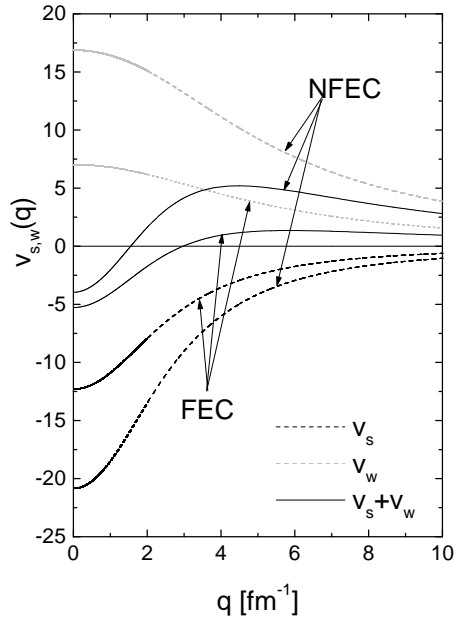}
\caption{The pairing potential and its repulsive and
attractive components in terms of the particle density, for FEC
(solid lines) and NFEC (dashed lines) approximations.}\label{FIG3}
\end{figure}

In Fig. \ref{FIG2} we display $\Delta(p_F)$ as a function of the
density, here one can see that the FEC treatment not only lowers
$\Delta_{max}$, but also reduces more than 20\% the range of
densities where the gap is effective. In this case $p_F=1.6$
fm$^{-1}$ is the upmost value for non-zero pairing gap.\\
 To improve understanding of this outcome, we investigate
separately contributions of scalar and vector character to the gap
function. The  attractive ($v_s$) and repulsive ($v_w$) potentials
evaluated at the Fermi surface
\begin{eqnarray}
v_s(q)&=&-\frac{g_s^2}{8 E_F E(q)}\left[1 +\frac{4 M^2-2 (E_F-
E(q))^2 -m_s^2}{2 p_F q}
\;\ln\left(\frac{(p_F-q)^2+m_s^2}{(p_F+q)^2+m_s^2} \right)\right]\nonumber \\
v_w(q)&=& -\frac{g_w^2}{8}\frac{2 E_F E(q)-M^2}{p_F \,q\, E_F\,
E(q)}\; \ln\left(\frac{(p_F-q)^2+m_w^2}{(p_F+q)^2+m_w^2}
\right)\nonumber
\end{eqnarray}
\noindent where $E_F=\sqrt{p_F^2+M^2}$, have been defined in order
that Eq. (\ref{SNMgap}) can be simplified to
\[ \Delta_3(p_F)=\int_0^\infty\, dq \, g_3(q)\left[ v_s(q)+v_w(q)
\right]. \nonumber
\]
As can be seen in figure \ref{FIG3}, both $v_s$ and $v_w$ appears
diminished in the FEC case, with a stronger suppression of $v_s$
as compared to $v_w$ in the high momenta regime. Therefore $v_s +
v_w$ asymptotically goes to zero faster in FEC than in NEFC,
minimizing contributions from higher momenta.\\
On the other hand, for  $q< 1$ fm$^{-1}$ the attractive potential
is stronger than the
\noindent
repulsive one in FEC than in NFEC approaches.\\
There is a surprising numerical similitude between the values
described above for $\Delta_{max}(p_F)$ and some findings in
\cite{CAO_prc74}, where a sophisticated evaluation of in medium
effects over the $^1S_0$ pairing is presented, within the Brueckner
theory of nuclear matter.  The BCS gap function is found to have a
maximum value  $\Delta_{max} \simeq 5$ Mev in symmetric nuclear
matter at $p_F=0.9 \mbox{fm}^{-1}$.  The upper limit
for the existence of the gap is given by $p_F=1.5$ fm$^{-1}$.\\
Another interesting comparison can be made with the results found by
Matsuzaki et al. \cite{MATSUZAKI2}. That work, as the present
approach, evaluate the possibility of giving a unified description
of both particle-particle and particle-hole channels. In that case
the sigma-omega model is used in the mean field approximation, and
the results are rendered physically acceptable by a direct
intervention over the integrals in momentum space. This is achieved
by the introduction of a form factor depending on a single
parameter, which is adjusted to obtain the best fit to microscopic
calculations. It must be pointed out that the correction factor
$\Theta$ used in our procedure can not be rigorously considered as a
form factor, since for a given Fermi momentum it reduces to a
constant value and therefore it does not modify the integrals.
Despite the procedural differences, our results are comparable to
those of \cite{MATSUZAKI2}. For instance, if we consider the
momentum dependence of the single-particle potential
$v(p_F,q)=v_s+v_w$, evaluated at $p_F$ such that a maximum of
$\Delta_{max}(p_F)$ is obtained, it takes values -4 MeV $< v(p_F,q)
<$ 3 MeV in \cite{MATSUZAKI2}. In our calculations, instead, we
found -5 MeV $< v(p_F,q) <$ 1 MeV, see Fig. 3. Furthermore, the
asymptotic behavior is similar in both calculations. More
appreciable differences are found for the momentum dependence of the
gap corresponding to the same $p_F$. At very low momentum we have
$\Delta \sim 6$ MeV, then as the transfer momentum $q$ is increased,
the gap decreases smoothly, passes through zero at $q \sim 3.5$
fm$^{-1}$, reaches a minimum value of -1 MeV at $q \sim 5.5$
fm$^{-1}$, and finally tends asymptotically to zero from negative
values, see Fig. 1. As can be seen in Fig. 3b of \cite{MATSUZAKI2},
the low momentum gap is sensibly lower $\Delta \sim 4$ MeV, the gap
passes trough zero at a lower value $q \sim 2$ fm$^{-1}$ and reaches
the same minimum value but at $q \sim 3.5$ fm$^{-1}$. From this
observations, we can conclude that in our approach, the
contributions coming from momenta $q < 3$ MeV are overestimated as
compared with the treatment of \cite{MATSUZAKI2}. Consequently, the
behavior of $\Delta_{max}(p_F)$ has a maximum value that exceeds by
2.5 MeV the results shown in Fig. 2a of \cite{MATSUZAKI2}. However,
this maximum value is reached for $p_F \sim 0.8$ fm$^{-1}$ in both
cases. The spreading is also similar, a drop of about 75$\%$ is
verified at $p_F \sim 1.2$ fm$^{-1}$ in both calculations.\\

\begin{figure}[ht]
\centering
\includegraphics[height=0.45\textheight]{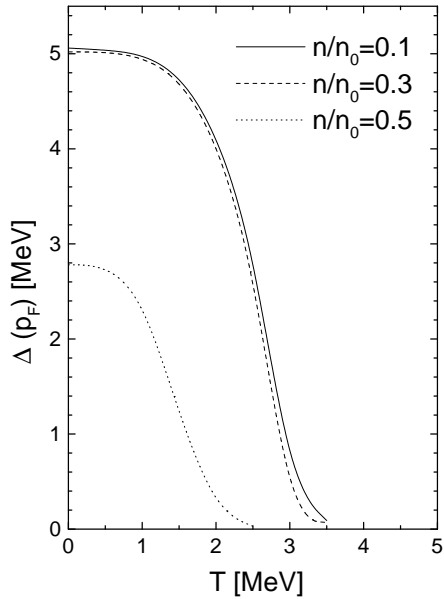}
 \caption{The gap energy for the $^1S_0$ phase in terms of the temperature for several
 densities.}\label{FIG4}
\end{figure}

As a last application we study the temperature behavior of the gap
function evaluated at the Fermi momentum in the FEC approach. In
fig. \ref{FIG4} we select some definite values of the particle
density such that the gap has magnitude higher than 0.1 MeV at
zero temperature. For the lower densities a steep fall is
registered around $T\sim 2.5$ MeV.

\section{Summary}

In this work we have studied the superfluid phase in a nuclear
environment, within the relativistic field theory of hadrons. The
Fermi-Landau liquid framework, as stated by \cite{AKHIEZER}, has
been used in order to obtain expressions for the $^1S_0$
superfluid gap energy and the distribution functions for both
normal and superfluid phases. This is equivalent to a Hartree
approximation, which provides a quasi-particle picture of nucleons
dressed by the mesonic interaction,  solved in a self-consistent
way with a BCS scheme for the superfluid phase.\\
It is well known that BCS approaches in terms of QHD models,
produce excessively large values for the gap in isospin symmetric
nuclear matter. In this work we explored the possibility to state
a easy to handle formalism, capable to retain the good properties
of the QHD theory but modifying the above mentioned failure. With
this purpose we took the simplest sigma-omega interaction and we
introduced short range correlations inspired by
 the finite volume extension of nucleons. In our approach an
 additional model parameter is required, which is related to the
 characteristic length of the finite size of nucleons. \\
 We have found a significative reduction, between $30-40 \%$, of the
 $^1S_0$ gap in symmetric matter. So, the results obtained are closer to
 non-relativistic estimates. The modifications proposed modify the high
 momenta performance producing a pairing potential which goes to
 zero faster than in \cite{KUCHAREK}. Further refinements, such
 as Fock term, additional mesons, non-linear sigma terms, etc., could
 improve the agreement with expected values.  \\
 Finally, the temperature behavior of the superfluid
 gap obtained is comparable with previous estimates \cite{MUTHER_prc72}, major
 differences correspond to the lowest densities.

\section*{Acknowledgements} This work was partially supported by the
CONICET, Argentina.


\begin{thebibliography}{99}
\bibitem{TANIGAWA}
T. Tanigawa, M. Matsuzaki, and S. Chiba, {\it Phys. Rev. C} {\bf
70} (2004) 065801.

\bibitem{SERRA}
M. Serra, A. Rummel, and P. Ring, {\it Phys. Rev. C} {\bf65}
(2001) 014304.

\bibitem{LONG_nt0812}
W. Long, P. Ring, N. Van Giai, and J. Meng, {\it Phys. Rev. C}
{\bf 81} (2010) 024308.

\bibitem{MUTHER_prc72}
H. Muther and W.H. Dickhoff,  {\it Phys. Rev. C} {\bf 72} (2005)
054313.

\bibitem{ZUO_prc66}
W. Zuo, U. Lombardo, H. Schulze, and C.W. Shen, {\it Phys. Rev. C}
{\bf 66} (2002) 037303.

\bibitem{GANDOLFI_prl101}
S. Gandolfi, et al.,  {\it Phys. Rev. Lett.} {\bf 101} (2008)
132501.

\bibitem{CAO_prc74}
L.G. Cao, U. Lombardo, and P. Schuck, {\it Phys. Rev. C} {\bf 74}
(2006)  064301.

\bibitem{MATERA} F. Matera, G. Fabbri, and A. Dellafiore, {\it Phys. Rev. C} {\bf
56} (1997) 228.

\bibitem{KUCHAREK}
H. Kucharek and P. Ring, {\it Z. Phys. A} {\bf 339} (1991) 23.

\bibitem{CHEN_plb585}
J.-S. Chen, P.-F. Zhuang, and J.-L. Li,{\it Phys. Lett. B}  {\bf
585} (2004) 85.

\bibitem{LiSunMeng}
J. Li, B. Y. Sun, and J. Meng, {\it Int. J. Mod. Phys. E} {\bf 17}
(2008) 1441.

\bibitem{MATSUZAKI3}
M. Matsuzaki, {\it Prog. Theor. Phys.} {\bf 116} (2006) 127.

\bibitem{DEAN_rmp}
D. J. Dean and  M. Hjorth-Jensen, {\it Rev. Mod. Phys.} {\bf 75}
(2003) 607.

\bibitem{MATSUZAKI0}
M. Matsuzaki, {\it Phys. Rev. C} {\bf 58} (1998) 3407

\bibitem{MATSUZAKI1}
M. Matsuzaki and T. Tanigawa, {\it Phys. Lett. B} {\bf 445} (1999)
254.

\bibitem{MATSUZAKI2}
M. Matsuzaki and T. Tanigawa, {\it Nucl. Phys. A} {\bf 683} (2001)
406.

\bibitem{QHD1a}
B. D. Serot and J. D. Walecka, {\it Advan. Nucl. Phys.} {\bf 16}
(1986) 1.

\bibitem{QHD1b}
B. D. Serot and J. D. Walecka,  {\it Int. J. Mod. Phys. E } {\bf 6}
(1997) 515.

\bibitem{HOFMANN_prc64}
F. Hofmann, C. M. Keil, and H. Lenske, {\it Phys. Rev. C } {\bf
64} (2001)  034314, and references therein.




\bibitem{StoGre}
D. H. Rischke, M. I. Gorenstein, H. St\"{o}cker and
                 W. Greiner, {\it Z. Phys. C} {\bf 51} (1991) 485.

\bibitem{Cley1}
J. Cleymans and H. Satz,  {\it Z. Phys. C} {\bf 57} (1993) 135.
\bibitem{Cley2}
H. Kouno, K. Koide, T. Mitsumori,
               N. Noda, A. Hasegawa, M. Nakano, {\it Prog. Theor. Phys.} {\bf 96} (1996)
               191.

\bibitem{Cley3}
               G. D. Yen, M. I. Gorenstein, W. Greiner, S. N. Yang, {\it Phys. Rev. C}
               {\bf 56} (1997) 2210.

\bibitem{Cley4} M. I. Gorenstein, H. St\"{o}cker, G. D.
               Yen, S. N. Yang, W. Greiner, {\it J. Phys. G} {\bf 24} (1998) 1777.

\bibitem{KAGIYAMA1}
S. Kagiyama, A. Nakamura, T. Omodaka,
               {\it Z. Phys. C} {\bf 53} (1992) 163.

\bibitem{KAGIYAMA2}
S. Kagiyama, A. Nakamura, T. Omodaka, {\it Z. Phys. C} {\bf 56}
(1992) 557.


\bibitem{KAGIYAMA3} S. Kagiyama, A. Minaka, A. Nakamura, {\it Prog. Theor.
               Phys.} {\bf 89} (1993) 1227.

\bibitem{SINGH1}
C. P. Singh, B. K. Patra, K. K. Singh, {\it Phys. Lett.}
                {\bf B387} (1996), 680.

\bibitem{SINGH2} R. Aguirre and A.L. De Paoli, LANL Report
{\it nucl-th/9907087}

\bibitem{SINGH3} P.K. Panda, M.E. Bracco, M. Chiapparini, E.
Conte and G. Krein,  {\it Phys. Rev. C} {\bf 65} (2002) 065206 .

\bibitem{KAPUSTA}
J. Kapusta, {\it Phys. Rev. D} {\bf  23} (1981) 2444.

\bibitem{AKHIEZER}
A.I. Akhiezer, V.V. Krasil'nikov, S.V. Peletminskii, and A.A.
Yatsenko, {\it Phys. Rep.} {\bf 245} (1994) 1.

\bibitem{ZAHED}
 I. Zahed and G. E. Brown, {\it Phys. Rep.} {\bf 142}  (1986) 1.

\bibitem{THOMAS}
J. Rikovska-Stone, P. A. M. Guichon, H. H. Matevosyan, and A. W.
Thomas, {\it Nucl.Phys. A}{\bf 792}(2007) 341.


\bibitem{ZIMANYI}
J. Zimanyi and S. A. Moszkowski, {\it Phys. Rev. C} {\bf 42}
(1990) 1416 .
\end{thebibliography}
\end{document}